# *DiSH* SIMULATOR: CAPTURING DYNAMICS OF CELLULAR SIGNALING WITH HETEROGENEOUS KNOWLEDGE

Khaled Sayed
Electrical and Computer Engineering
University of Pittsburgh
3700 O'Hara Street, Pittsburgh PA 15261, USA
k.sayed@pitt.edu

Yu-Hsin Kuo
Language Technologies Institute
Carnegie Mellon University
5000 Forbes Ave, Pittsburgh PA 15213, USA
yhsinkuo@gmail.com

Anuva Kulkarni
Electrical and Computer Engineering
Carnegie Mellon University
5000 Forbes Ave, Pittsburgh PA 15213, USA
anuvak@andrew.cmu.edu

Natasa Miskov-Zivanov
Electrical and Computer Engineering, Bioengineering,
Computational and Systems Biology
University of Pittsburgh
3700 O'Hara Street, Pittsburgh PA 15261
nmzivanov@pitt.edu

**ABSTRACT**

We present DiSH-Sim, a simulator for large discrete models of biological signal transduction pathways, capable of simulating networks with multi-valued elements in both deterministic and stochastic manner. We focus on order of update and thus incorporate information about timing, taking into account that biological processes are not synchronized and certain biochemical changes occur slower than others. Another feature of our simulator is the use of grouped rules to model multi-valued elements and delays. The DiSH-Sim is publicly available and is being used to validate discrete cancer microenvironment and infectious disease models. It is also incorporated within a large architecture that includes natural language processing tools that read biological literature to assemble logical models. This paper demonstrates the functionalities and ease of use of DiSH-Sim, making it a very useful tool for discrete modeling.

## 1 INTRODUCTION

Biological signaling pathways consist of molecules that interact with each other to convey signals inside and outside the cell (Gomperts et al., 2009). The pathways and their components can be modeled as graphs, where nodes represent pathway components (e.g., proteins, genes, chemicals, or even cellular processes), and edges represent component interactions (e.g., biochemical reactions). The graphs illustrate qualitative information about systems, reveal that a particular causal relationship exists, or that a certain node positively or negatively regulates its neighbor. Feedback and feed-forward loops are also easily visualized using graphs.

To study the *behavior* of pathway components and overall systems *in time*, as a response to external signals or internal perturbations, *executable models* are built and simulated. Executable models implement graph nodes as variables, and interactions as variable update functions. Models may also contain quantitative information about reaction rates and component concentrations. The simulations start with specified initial conditions that represent particular pathway or cell state, and continue by executing update functions of model elements until desired state is observed, or until the model reaches steady state.

Executable models of cellular signaling pathways can be either continuous (Karlebach et al. 2008; Sobie et al. 2011) or discrete (Pinney et al. 2003; Thomas et al. 1990). Continuous models (e.g., ODE-based models) contain quantitative information such as concentrations of reactants and products and reaction rates. They are used when the system being modeled is well understood through experimental observations. However, the information about mechanisms of direct interactions and their parameters is often not available, or the modeler might prefer not to use all of the information available due to computational limitations. In these cases, discrete modeling approach can be used.

It has been shown that Boolean models, a special case of discrete models, are capable of capturing characteristic dynamic behavior such as multi-stability, excitation and adaptation behavior (Albert et al., 2015). Positive and negative feedback and feed-forward loops can also be implemented in these models (Miskov-Zivanov et al., 2013). Incomplete information can be dealt with by using indirect causal evidence, which is not possible in ODE models or reaction rule-based models (Faeder et al., 2009). In discrete models, each element is assigned a set of discrete values representing activity of the modeled system component, as well as an update rule, which is a function of its regulators. In discrete models, update rules change values of elements over discrete simulation time steps.

Dynamic model analysis approaches have been developed previously (Danos et al., 2008; Faeder et al., 2009; Gillespie, 1977; Kochańczyk et al., 2017; Varga et al., 2017). Several of these tools are used particularly for simulating Boolean models and some packages allow multi-valued modeling as well (Di Cara et al., 2007; Gonzalez et al.,2006; Hinkelmann et al., 2011). Additionally, models can involve both intra-cellular and extra-cellular processes, and hence it is necessary to build a tool to study the dynamics of complex but coarse-grained networks, in order to understand their behavior and response to stimuli. The *timing* of component changes is an important consideration in biological simulations. Different biological processes and reactions taking place inside a cell and in the cells environment are not synchronized and can even happen at different time scales. Furthermore, some components and pathways in modeled systems are well studied such that their mechanistic details are available, while other parts of the system are less known or not well understood. Finally, the speed of simulations remains critical, especially when simulation is a part of a larger framework. One such recent example includes automated information extraction, assembly of many model variants, model selection and analysis, with a feedback to guide further information extraction when needed, for the purpose of system understanding, explanation and prediction(Cohen, 2015; Miskov-Zivanov, 2015). Discrete modeling approach has been shown to be advantageous in all these situations (Miskov-Zivanov et al., 2014).

In this paper, we present our simulator, DiSH (Discrete, Stochastic, Heterogeneous model simulation), a stochastic simulator for discrete models with heterogeneous elements and interactions. DiSH implements multiple simulation schemes, and allows for fast simulation of discrete models.

Compared to other existing simulators (I. Albert et al., 2008; Bock et al., 2014; Danos et al., 2008; Faeder et al., 2009; Gonzalez et al., 2006; Helikar et al., 2009; Müssel et al., 2010; Yu, Tung et al., 2012; Zheng et al., 2010), our contribution with DiSH is two-fold: (1) we developed a representation that allows for multi-valued hierarchical models to be translated into logic circuit-like models and used by our simulator, and (2) we incorporated rates of biological processes by associating probabilities with element update rules. Section 2 provides background on discrete modeling and Section 3 describes different simulation schemes that are supported by our simulators. Section 4 compares and contrasts the simulation schemes using the model from (Miskov-Zivanov et al., 2013) and Section 5 concludes the paper.

## 2 BACKGROUND

The construction of a model begins with identifying key signaling pathways of the modeled system, and the components on these pathways. Figure 1(a) outlines a modeling framework that uses our DiSH simulator. The framework includes identifying the key signaling pathways through literature reading, data analysis, and/or expert knowledge. The information from these different sources is then translated to a common representation formalism, from which executable models are created. Additionally, the elements in the executable models have to be initialized before running simulations to implement a particular system state. Finally, after simulations, the results can be visualized in two ways, using plots that show transient behavior and steady-state for each element, or using summaries of element behavior accross different initial states and assuming different system inputs. The feedback loop from simulation results to information sources allows for further targeted information extraction from sources.

Formally, we can describe discrete models by *model elements* and *influence sets*. Let $\mathcal{E}$ be a set of all *model elements* (i.e., graph nodes), $\mathbf{E}_i \in \mathcal{E}$, i=1,..,N, N=|$\mathcal{E}$|, that represent these system components. For each model element, we determine its *influence set*, which includes the element's positive and negative regulators, also called activators and inhibitors, respectively. The influence sets can be illustrated as *interaction maps* (graphs), where nodes represent model elements and edges represent regulatory interactions between elements. In Figure 1(b), we show an example interaction map of the CD4+ T-cell model from (Miskov-Zivanov, Turner, et al., 2013). We also determine the *number of values*, $v_i$, that each element, $E_i$, can take. These values represent the number of *discrete states* that a system component is observed to have.

Next, we assign a set of Boolean variables $E_{i,m}$, and their corresponding update functions $f_{E_{i,m}}$ where $m=0,..,M_i$ and $M_i = (\log_2 v_i)-1$ to each element $E_i$. The variables and their update rules comprise *executable model* that is used by our simulator. Elements often have only two states: high activity or concentration (1 or True or ON value of Boolean variable) and low activity or concentration (0 or False or OFF value of Boolean variable). In this case, elements are assigned a single Boolean variable. However, model elements are not restricted to Boolean states, and can have multiple states that are then encoded with multiple Boolean variables. This allows modelers to encode multiple levels relevant to element's downstream activity, rather than encoding only two levels, while still utilizing logical model simulators and their benefits (Miskov-Zivanov & Kashinkunti, et al., 2011; Miskov-Zivanov & Marculescu, et al., 2011). For example, element PI3K from the T-cell model in (Miskov-Zivanov et al., 2013) has three states, representing the three observed levels of interest of PI3K in T cells, namely LOW, MEDIUM and HIGH. Therefore, PI3K element from the influence map in Figure 1(a) is represented in the executable model using two variables $PI3K_1$ and $PI3K_0$. Note that the combinations of variable $PI3K_1$ and $PI3K_0$ values are 00, 01, 10 and 11, where only three of these combinations are needed to represent the three states of the PI3K node:

$$PI3K = \begin{cases} \text{LOW}, & (PI3K_1, PI3K_0) = 00 \\ \text{MEDIUM}, & (PI3K_1, PI3K_0) = 01 \\ \text{HIGH}, & (PI3K_1, PI3K_0) = 10 \end{cases} \quad (1)$$

This will be the case whenever $\log_2$ of the number of levels modeled is not an integer. In such cases, we will round $\log_2$ of the number of levels to the next integer. The additional variable value combinations obtained due to rounding (($PI3K_1,PI3K_0$) = 11 in the case of PI3K) are either prohibited by the construction of the model, or made redundant by assigning them to one of the existing states. For example, since PI3K has only three states (LOW, MEDIUM and HIGH), the combination ($PI3K_1,PI3K_0$) = 11 will never occur in the model in (Miskov-Zivanov et al., 2013).

Furthermore, the number of discrete values in regulated element and the number of discrete values in its regulators do not have to be same. For example, in (7)(a), the binary-valued element mTORC1 can be active or inactive based on the values of its regulators, PI3K and S6K1. Note that S6K1 is also a binary-valued element, and therefore it requires a single Boolean variable, while PI3K is encoded with two Boolean variables, $PI3K_1$ and $PI3K_0$, each one having its own update rule.

More formally, in general, the value of multi-valued element $E_i$ is computed as:

$$E_i = E_{i,0} \cdot 2^0 + \cdots + E_{i,M_i} \cdot 2^{M_i}. \quad (2)$$

The next value of variable $E_{i,m}$, at time step $t+1$, $E_{i,m}^{t+1}$ can be computed as a function of the regulators of element $E_i$ at time $t$, that is, a function of Boolean variables corresponding to the regulators of element $E_i$:

$$E_{i,m}^{t+1} = f_{E_{i,m}}\left(E_{1,0}^t, \ldots, E_{1,M_1}^t, E_{2,0}^t, \ldots, E_{2,M_2}^t, \ldots, E_{N,0}^t, \ldots, E_{N,M_N}^t\right). \quad (3)$$

We also define here a function $f_{E_i}^t$, representing the value that can be computed from regulators of element $E_i$ at time step $t$:

$$E_i^{t+1} = f_{E_i}^t = f_{E_i,0}^t \cdot 2^0 + \cdots + f_{E_i,M_i}^t \cdot 2^{M_i} \quad (4)$$

In Figure 1(c), we show a toy example of interaction map that we will use throughout the paper to explain features of our simulator. The example includes three elements, that is, their corresponding Boolean variables and, for each variable, a logic update rule from the executable model. Multiple regulators can be combined in element update rule to compute the next state of element. The update rule is composed of variables corresponding to element's regulators and of logic functions AND, OR and NOT. A change in the state of regulators is reflected in the change of the regulated element, according to the logic combining regulators in the update rule – negatively (NOT), independently (OR), or with conditional dependence (AND). The creation of logic rules is described in (Miskov-Zivanov et al., 2013). We have developed software which can derive logic update rules for all model variables according to influence set notation (determine when to use logic operators AND, OR, NOT). The details of the notation are beyond the scope of this paper. Finally, given that logical models do not represent the time of biological events in the same manner as ODE models, methods exist to incorporate timing into these models (Miskov-Zivanov et al.,

2014). Our simulation schemes described in the next section are designed to account for these delay modeling methods.

## 3 SIMULATION SCHEMES

In this section, we describe model execution schemes that our simulator supports (also shown in Figure 1(d)). Two main categories of simulation schemes are supported by DiSH: *Simultaneous* (Sim) or synchronous, and *Sequential (*Seq*)* or asynchronous scheme. As shown in Figure 1(d), in the Seq scheme, elements can be updated using either *Ranked-order* (RankSeq) or *Random-order* (RandSeq) scheme. In addition, the selection of an element to be updated next in the RandSeq scheme can be done in two ways: *Round-based (*RB-RandSeq*)* selection and *Step-based (*SB-RandSeq*)* selection, as described in the subsequent sections. The probability of selecting a rule to be updated in the SB RandSeq scheme can be *Uniform* (USB-RandSeq), that is, the same probability value is assigned to each logic rule, or *Non-Uniform* (NUSB-RandSeq), where the assigned probability is different for each rule.

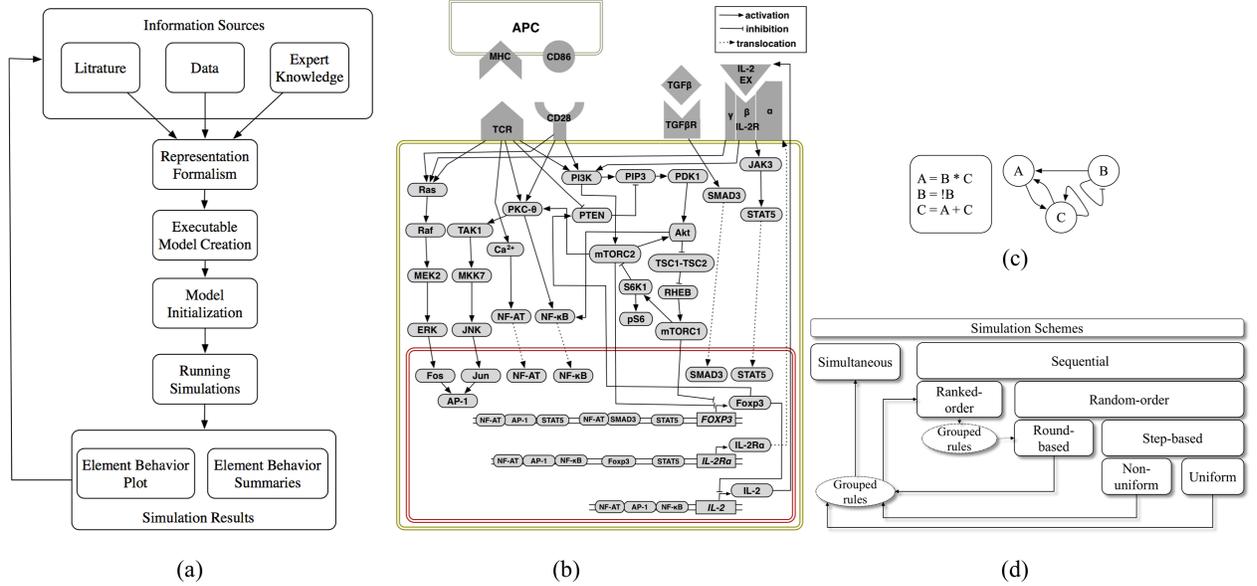

(a) (b) (d)

Figure 1: (a) Modeling flow. (b) An interaction map of the T-cell model from (Miskov-Zivanov, Turner, et al., 2013). (c) A toy example: three nodes (A, B and C), and their update rules specified (here, logic rules, * represents AND logic operator, + represents OR logic operator, and ! represents NOT logic operator). (d) Simulation schemes.

Formally, the following holds for all the simulations schemes that we use. Let § be the set of all possible states of the system, then we can compute the size of § as:

$$|§| = v_1 \cdot v_2 \cdot \ldots \cdot v_N \quad (5)$$

where $v_i$ is the number of discrete values that an element $E_i \in \mathcal{E}$ can take, and $N$ is the size of set $\mathcal{E}$, as defined above. In general, the probability $p(S^{t+1}=S_k)$ that, at time step $t+1$, the state $S$ of the model is equal $S_k$ where $k$ is an integer between 1 and $|§|$, can be computed as:

$$p(S^{t+1} = S_k) = \sum_{l=1}^{|S|} p(S^{t+1} = S_k | S^t = l) \cdot p(S^t = S_l) \quad (6)$$

Furthermore, the probability that element $\mathbf{E}_i$ will take value $V_j$ where $j=0,..,v_i-1$ in time step $t+1$ during simulation, $p(\mathbf{E}_i^{t+1} = V_j)$, is then given by the following equation:

$$p(\mathbf{E}_i^{t+1} = V_j) = P_{\mathbf{E}_i} \cdot p(f_{\mathbf{E}_i}^t = V_j) + (1 - P_{\mathbf{E}_i}) \cdot p(\mathbf{E}_i^t = V_j) \quad (7)$$

where $P_{\mathbf{E}_i}$ represents the probability that element $\mathbf{E}_i$ is selected to be updated next, and $f_{\mathbf{E}_i}^t$ is the value of the update function for element $\mathbf{E}_i$ at time step $t$ during simulation. Then, $p(f_{\mathbf{E}_i}^t = V_j)$ can be computed as:

$$p(f_{E_i}^t = V_j) = \sum_{l=1}^{|S|} p(f_{E_i}^t = V_j | S^t = S_l) \cdot p(S^t = S_l) \tag{8}$$

Therefore, one can then compute the probability of state $S^{t+1}$ being equal to state $S_k = (V_{k,1}, V_{k,2}, ..., V_{k,N})$, where $V_{k,i}$ is the value of element $E_i$ in state $S_k$, as:

$$p(S^{t+1} = S_k) = \sum_{i=1}^{N} p(E_i^{t+1} = V_{k,i}) \tag{9}$$

### 3.1 Simultaneous Scheme

In the Sim simulation scheme, all elements are updated simultaneously, that is, current state values of all variables are used to simultaneously compute next state values. This simulation scheme is therefore deterministic: for each state, there is only one possible next state. In this simulation scheme, if an initial state of the system is given, one can determine the steady state or steady cycle that the system will reach. In other words, the probability $P_{E_i}$ from equation (7) is equal 1 for all elements $E_i$, and in each given time step $t$, the probabilities $p(S^t = S_l)$ will be equal 0 for all but one value $l=L$, for which it will be $p(S^t = S_L) = 1$.

A state transition graph (STG) resulting from the simulation of our toy example from Figure 1(d), using the Sim scheme, is depicted in Figure 2(a). Given elements of the toy model, A, B, C, and their corresponding values, $V_A$, $V_B$, $V_C$, if the simulation starts at time $t=0$ we denote this as state $S^0 = (V_A^0, V_B^0, V_C^0)$. The probability of the following states $S^{t+1} = S_k$ for time steps $t=0,...,n$ where $n$ is the number of simulation steps, will in each time step be equal 1 for one specific $k=K_t$, and will be 0 for all other $k \neq K_t$. For example, in our toy system, as shown in Figure 2(a), if the simulation starts from any state except states 000 and 010, we will always reach state 101, and the model will then oscillate between states 011 and 101. In addition, if the simulation starts at state 000 or 010, it keeps oscillating between these two states and never moves to any other state.

### 3.2 Random-order Sequential Scheme

In the RandSeq scheme, model variables are not updated simultaneously, instead they are updated sequentially in random order. In other words, once element $E_i$ is updated in time step $t$ by computing its new value according to its update function $f_{E_i}$, this new value is used to determine model element values in the following time steps until the same element $E_i$ is selected for update again. This scheme allows for modelling cellular signaling and processes in a more realistic way than in the Sim scheme, as it accounts for the randomness that exists in the timing of biological events. The STG of our toy example system, when simulated using the RandSeq scheme, is shown in Figure 2(b). As can be seen from the STG, a state can have multiple next states, and thus, a given initial state can be followed by multiple different paths through STG and result in different steady-states. This means that the probability $P_{E_i}$ from equation (7) does not have to be equal 1 for a given element $E_i$ in a given time step $t$, as it was the case in the Sim scheme. Instead, this probability will depend on the method used for selecting elements for update, as described in the following subsections. Furthermore, in each given time step $t$, the probabilities $p(S^t = S_l)$ can vary between 0 and 1 for different values of $l$.

### 3.2.1 Round-Based Random-order Sequential Scheme

The RB-RandSeq simulation scheme has been previously described in (I. Albert et al., 2008; Li, Assmann, & Albert, 2006) and used in (Miskov-Zivanov, Turner, et al., 2013). It is important to distinguish here between simulation *step* and simulation *round*. While the simulation step accounts for updating value of a single element, and can also correspond to time step in our earlier discussion, the simulation round represents a cycle within which **all** elements are updated **exactly once** according to their update functions. Formally, if a model has $N$ elements, $E_i$, $i=1,...,N$, and their update functions are $f_{E_i}$, then each round consists of $N$ steps, formally ROUND = (STEP$_1$,...,STEP$_N$), and in each round a new random order is determined, in which the values of these function are computed. Thus, in a given round $T$, the element update order is a random permutation $\mathcal{P}_r$ of the vector (STEP$_1$,...,STEP$_N$):

$$\text{update\_order}_T (\mathcal{E}) = (\text{STEP}_{T1},...,\text{STEP}_{TN}) = \mathcal{P}_r(\text{STEP}_1,...,\text{STEP}_N) \tag{10}$$

such that the step at which element $\mathbf{E}_i$ is updated is $STEP_{T_i}$. Given that every element gets updated exactly once within a round, the probability $P_{\mathbf{E}_i}$ that an element $\mathbf{E}_i$ is selected for update in a given time step $t$ depends on the $STEP_{T_i}$ within round when this update occurs:

$$P_{\mathbf{E}_i} = \frac{1}{N - (STEP_{T_i} - 1)} \quad (11)$$

The two simulation rounds when the RB-RandSeq method is applied to our toy example starting at state 100 are shown in Figure 2(c).

### 3.2.2 Step-Based Random-order Sequential Scheme

In the SB-RandSeq simulation scheme, one model element is chosen for update in each time step. There are no round-based restrictions for element selection, and therefore, the same element can be updated in consecutive steps. In addition, elements can be chosen for update all at the same rate or at different rates. Thus, we define two sub-schemes:

*a) Uniform (*USB-RandSeq*):* In this simulation approach, all elements have the same update rate, and for a model with *N* elements, in each step the probability for an element $\mathbf{E}_i$ to get selected for update is:

$$P_{\mathbf{E}_i} = \frac{1}{N} \quad (12)$$

This approach is used when the time scales of changes in system elements are not well known, and thus the default approach is to assume that the rates at which elements are updated are equal.

*b) Non-uniform (*NUSB-RandSeq*):* In this simulation approach, each model element $\mathbf{E}i$ has an assigned update rate, $r_{\mathbf{E}_i}$. We assign update rates to elements based on prior knowledge, such that the system evolves over time following the rate of change observed in experiments. For example, since gene transcription and translation occur at a different time scale compared to protein modifications, we assign higher rates to protein interactions and lower rates to gene activation and protein synthesis. In this simulation approach, the probability of element being selected next for update is a function of these update rates:

$$P_{\mathbf{E}_i} = \frac{r_{\mathbf{E}_i}}{\sum_{i=1}^{N} r_{\mathbf{E}_i}} \quad (13)$$

where the probability of selecting an element $\mathbf{E}i$ to update its rule is proportional to the sum of all update rates in the model. We illustrate both types of the SB scheme, USB and NUSB, in Figure 3(a) and (b).

### 3.3 Ranked

Element update rules can also be assigned *rank numbers*. This feature is adopted from the BooleanNet tool developed by (I. Albert et al., 2008). Those rules that have same rank are executed using RB-RandSeq scheme. Similarly, groups with different rank are executed according to their rank: all rules in the group with rank 1 are executed first, then all the rules with rank 2 are executed, and so on. As shown in Figure 4(a)(left), B and C should be updated first in random order before we update A which has rank 2. Additional

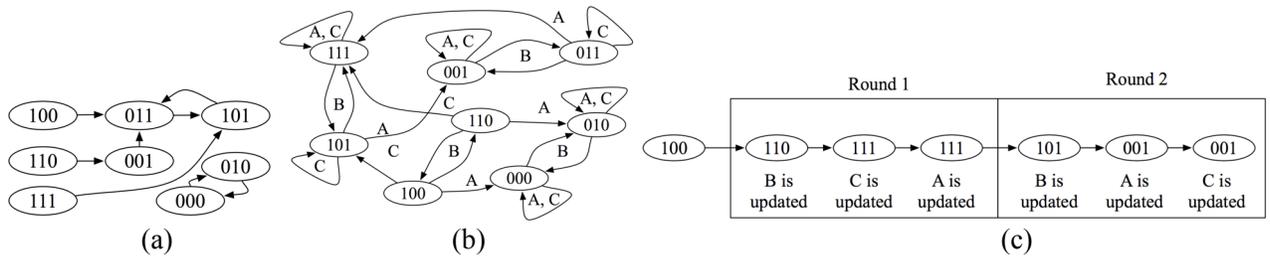

Figure 2. State transition graph (STG) for the toy example in Figure 1, for (a) the Sim scheme and (b) the RandSeq scheme. Labels on graph edges indicate which elements are selected for update when moving from the current state to the next state. (c) RB-RandSeq scheme.

functionalities that can be combined with the ranked method are described in the following section.

### 3.4 Additional Functionalities

Here, we describe additional features of the simulator that are either used as part of the simulation schemes described above, or represent ways to combine these schemes.

### 3.4.1 Grouped rules

There are several cases within sequential schemes in which it is required for variable update rules to be grouped together and computed simultaneously, or in the order they are listed in the model file: (i) all Boolean variables $E_{i,m}$ representing the *same model element* $\mathbf{E}_i$ are grouped and updated simultaneously; (ii) if there are *different model elements* that need to be updated at the same time, all their corresponding variables will be grouped and updated simultaneously; (iii) if it is required for a group of different model elements or for a group of model variables corresponding to the same element to be updated in a specific order, but in random order with the other elements or variables in the model, they are grouped and executed sequentially, in the order their update rules are listed in the model file.

The update rules of of all variables that need to be updated together are specified within curled braces '{}' in the model file. In synchronous scheme, this does not change the execution of rules, since the grouped rules are executed at the same time with the other rules. However, in the asynchronous scheme, the grouped rules are executed together when the group is selected to be updated at a specific time step. Figure 4(b) shows an example where nodes A and C are grouped. The difference in resulting simulated model behavior is illustrated with two state diagrams when the first (initial) state is the same, as shown in Figure 4(c).

### 3.4.2 Toggle implementation

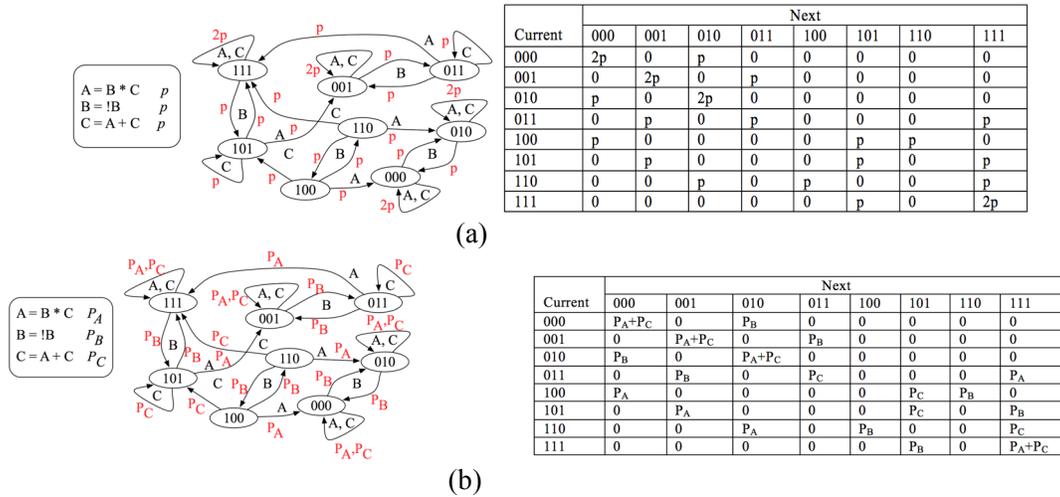

Figure 3. Logic rules (left), state transition graph (middle), and a table of the probability of change from one state to another (right) in the (a) USB-RandSeq and (b) NUSB-RandSeq Scheme

It is possible to toggle the value of a variable (i.e., switch from 1 to 0 and from 0 to 1) at a specific round or step by indicating it in the model file next to the variable initialization. This is often a useful feature of simulations that allows us to closely mimic wet-lab experiments. For example, toggling the value of the T-cell receptor (TCR) signal in the T-cell model allows us to study the impact of the duration of a high signal on the system's behavior. This functionality should be used with the RandSeq simulation schemes.

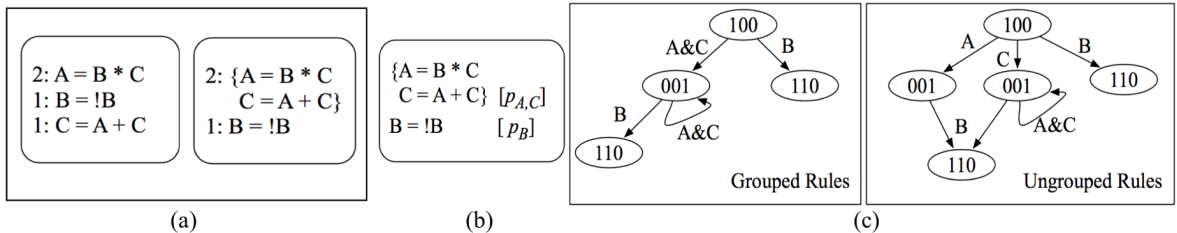

Figure 4. (a) (left) Ranked rules and (right) Ranked rules with grouping. (b) The syntax for grouped rules. (c) Two small examples of the state diagrams when grouped rules are used.

## 4 SIMULATION RESULTS

In the following discussion we will refer to <u>run</u> as a single simulation run from the starting point, when we assign initial values to all variables, through a pre-determined number of rounds or number of steps (depending on the simulations scheme used). A *trajectory* of values is obtained for each element in the run and the trajectories may vary across consecutive runs. The values for variables in each round or step are averaged over all the runs and this *average trajectory* can be plotted for analysis.

## 4.1 Case study – T cell differentiation model

In this section, we illustrate different simulation schemes that are implemented within our DiSH simulator using the T-cell differentiation model from (Miskov-Zivanov, Turner, et al., 2013). The stimulation of naïve peripheral T cells occurs via antigen presentation to T cell receptor (TCR) and with co-stimulation at CD28 receptor. These two stimulatory signals lead to activation of several pathways, of feedback and feedforward loops between pathway elements, and eventually result in differentiation of naïve T cells into helper (Th) or regulatory (Treg) phenotype. It has been shown that the ratio between Th and Treg cells in the T cell population strongly depends on antigen dose where high antigen dose results in mostly Th cells, while low antigen dose leads to mixed population of Th and Treg cells. Therefore, TCR is modeled in (Miskov-Zivanov, Turner, et al., 2013) as having three different values, no stimulation (value 0, or LOW according to equation (1)), stimulation with low antigen dose (value 1, or MEDIUM according to equation (1)), and stimulation with high antigen dose (value 2, or HIGH according to equation (1)). According to the description in Section 2, element TCR is implemented in the executable model using two Boolean variables, $TCR_{HIGH}$ and $TCR_{LOW}$. Furthermore, the duration of the presence of the high signal (high antigen dose, $TCR_{HIGH}=1$) has been shown to be critical in phenotype decision (Miskov-Zivanov, Turner, et al., 2013).

Here, we simulate two main scenarios; Scenario I, which includes 5 sub-scenarios, and Scenario II, which includes 3 sub-scenarios, all of them listed in Table 1. The five sub-scenarios under Scenario I consider low and high levels of antigen dose ($TCR_{HIGH}=0$ and $TCR_{LOW}=1$, $TCR_{HIGH}=1$ and $TCR_{LOW}=0$, respectively), as well as different initial values of proteins TGF-β and AKT, also responsible for regulating the T-cell differentiation. It has been shown that the addition of TGF-β resists the suppression of Foxp3 under the condition of high antigen dose (Miskov-Zivanov, Turner, et al., 2013) while the removal of AKT induces the expression of Foxp3 and enhances the differentiation into Treg cells (Sauer et al., 2008). On the other hand, the three sub-scenarios of Scenario II consider the antigen dose removal at certain time steps to reflect the impact of the time at which the high antigen dose is removed during the wet-lab experiments on the differentiation outcomes of the naïve T cells. The removal of the high antigen dose is simulated by toggling the value of the $TCR_{HIGH}$ variable from 1 to 0 at specific simulation steps, which changes the value of the TCR variable from HIGH to LOW, according to equation (1).

Table 1: Simulation Scenarios Summary.

| Scenario name | Description | Specific Conditions |
| --- | --- | --- |
| I1 | High Antigen dose | $TCR_{HIGH}$ = True |
| I2 | Low Antigen Dose | $TCR_{LOW}$ = True |
| I3 | High antigen dose and TGF-β is high | $TCR_{HIGH}$= True, $TGF_{BETA}$ = True |
| I4 | Low antigen dose and TGF-β is high | $TCR_{LOW}$ = True, $TGF_{BETA}$ = True |
| I5 | High Antigen Dose and AKT is off | $TCR_{HIGH}$ = True, $AKT_{OFF}$ = True |
| II1 | High antigen dose, toggle $TCR_{HIGH}$ | Toggle at Step 100 |
| II2 | High antigen dose, toggle $TCR_{HIGH}$ | Toggle at Step 300 |
| II3 | High antigen dose, toggle $TCR_{HIGH}$ | Toggle at Step 500 |

## 4.2 Comparison between different simulations schemes

In the subsequent sections, we show the simulation results obtained by running the simulator using the simulation schemes that were described in Section 3, for Foxp3 only due to the limited space. However, we discuss the response of Foxp3 in terms of other key elements such as IL-2, mTORC1, CD25, and STAT5, for the scenarios shown in Table 1. These 5 model elements were selected to describe the outcomes of the naïve T cells differentiation process into Th and Treg cells, where Th cells are characterized by the high expression of IL-2 and low expression of Foxp3, while the Treg cells are characterized by the high expression of Foxp3 and low expression of IL-2 (Miskov-Zivanov, Turner, et al., 2013). In addition, mTORC1 is a key player in the inhibition of Foxp3 at high antigen dose, while the early activation of Foxp3 by CD25/STAT5 pathway is an essential requirement for the differentiation of the naïve T cells into Treg cells at low antigen dose (Miskov-Zivanov, Turner, et al., 2013).

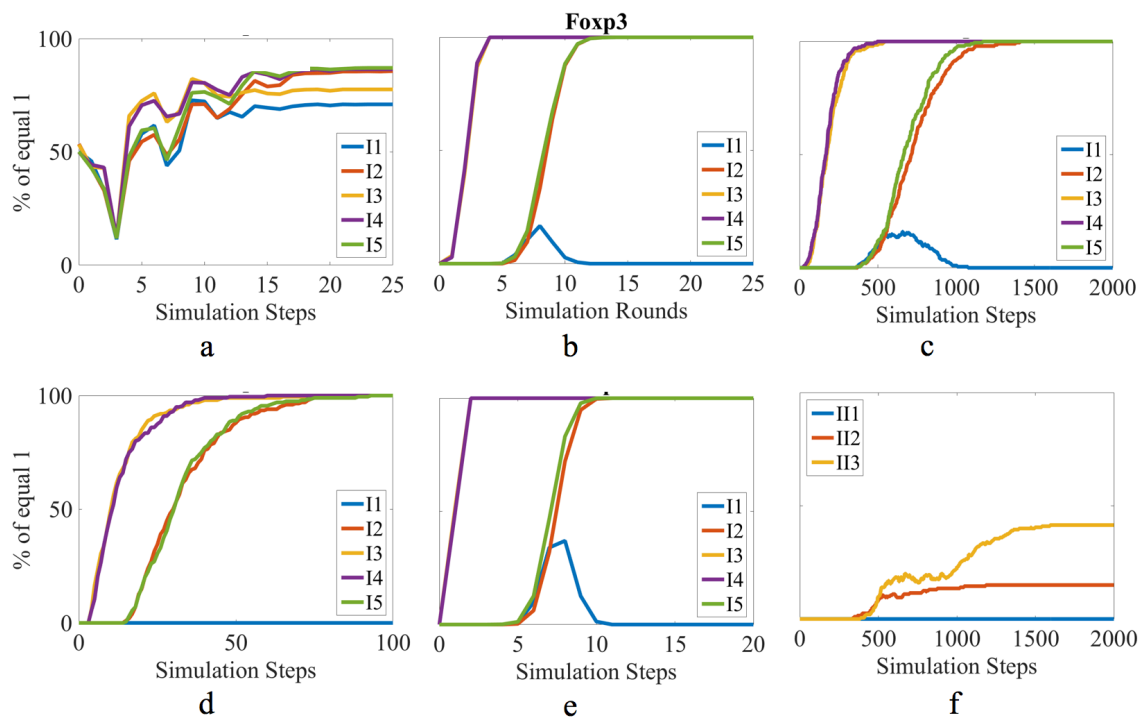

**Figure 5: Trajectories for Foxp3 using a)** Sim, b) RB-RandSeq, c) USB-RandSeq, d) NUSB-RandSeq, e) RankSeq , and f) Toggling simulation schemes

### 4.2.1 Simultaneous Scheme

The Sim scheme is deterministic as each state has only one possible next state. All variable values at time $t+1$ are computed using the values of their regulators at time $t$. This simulation scheme computes steady-states that the system can reach, but it cannot account for the randomness of occurrence of events that is common for biological systems.

In Figure 5(a), we show the simulation results for Scenarios I using the Sim scheme. The T cell model is simulated 1000 times (i.e., 1000 runs), each run consisting of 50 steps, assuming random initial values for all elements, except for TCR, TGF-β and AKT which were selected to satisfy the requirements of each sub-scenario in Table 1. It can be seen that the steady-state value of Foxp3 in sub-scenario I1 (high antigen dose) is lower than the steady-state value in sub-scenario I2 (low antigen dose), which was expected, according to (Miskov-Zivanov, Turner, et al., 2013) . However, the steady-state value of Foxp3 is not very low in sub-scenario I1 because the expression level of STAT5 and CD25 is high (activators of Foxp3), even with high expression level of mTORC1 (inhibitor of Foxp3). Also, we can see some oscillations in the transient response of Foxp3 in Figure 5(a), which arise from the random initializations of the model elements enabling each simulation run to start from a different state, i.e., a different point in the state space. Additionally, these plots show that the Sim scheme is useful for quickly identifying different steady-states that the system can reach. However, the Sim scheme may not provide accurate trajectories for studying the transient response of the system, as it does not account for the stochasticity in the biological systems.

### 4.2.2 Round-Based Random Sequential Scheme

A more detailed analysis can be performed in RandSeq scheme, which computes transient behavior of elements, and provides a better resolution of small changes occurring on element trajectories. As described in Section 3.2, the RandSeq scheme has two sub-types, round-based (RB-RandSeq) and step-based (SB-RandSeq) schemes. Here, we show simulation results using the RB-RandSeq scheme for all sub-scenarios of Scenario I, where the value of each element is updated once per round according to the element's update rule. The results of our DiSH simulator are in agreement with those presented in (Miskov-Zivanov, Turner, et al., 2013), that were obtained using the simulator described in (I. Albert et al., 2008). As shown in Figure 5(b), the steady-state value of Foxp3 is low (high) at scenario I1 (I2) which represents the high (low) antigen dose. In addition, Foxp3 exhibits a transient increase at high antigen dose (Scenario I1), due to the activation of CD25 and STAT5. This increase of the Foxp3 activation is quickly turned off because of the activation of mTORC1, which is a Foxp3 inhibitor. We can also see that initializing TGF-β at high level, with low antigen dose stimulation (Scenario I4) can provide a rapid increase in the Foxp3 expression, which inhibits any transient response of IL-2.

### 4.2.3 Uniform-Probabilities Step-Based Random Sequential Scheme

The other sub-type of the RandSeq simulation scheme is the Step-based scheme which has also two sub-types, the uniform (USB) and non-uniform (NUSB) probability simulation schemes. The simulation results for Scenario I with the USB-RandSeq scheme are shown in Figure 5(c). The simulation results in Figure 5(c) are similar to the ones inFigure 5(b), where the RB-RandSeq scheme was used, except that the transient responses shown in Figure 5(c) are delayed. This is happening due to the nature of the updating scheme. So, while each element gets updated once per round in the RB-RandSeq scheme, only one element is updated in a step in the SB-RandSeq scheme. Due to such updating schemes, the RB simulations will show faster rates of change than the SB ones. In the RB simulation, the number of time steps in each round is equal to the number of variables. In the T cell model, there are 61 variables, and in each of the 50 rounds, each variable is updated 50 times. On the other hand, in the step-based simulation, it may take more than 50 steps to update all elements because some elements may be updated several times within those 50 steps, while some other elements will not get updated. Figure 5(c) also emphasizes an interesting biological finding which was confirmed by the experimental results in (Miskov-Zivanov, Turner, et al., 2013), suggesting that initializing TGF-β at high even with high antigen dose stimulation (Scenarios I3 and I4) will produce more Treg cells.

### 4.2.4 Non-Uniform-Probabilities Step-Based Random Sequential Scheme

Here, we show the simulation results using the NUSB-RandSeq scheme, where in each time step, one variable is chosen for update according to the assigned update probabilities. When studying the T cell model using this scheme, we divide all the variables of the T-cell model into two blocks. Block A contains CD25, Foxp3 and IL2 variables, and has lower probability value, which is 0.1, and Block B which contains the rest of the variables with probability 0.9. The blocks have been constructed using prior knowledge about the biological system – it is known that protein-protein interactions (Block B) occur at a faster speed than transcription reactions in Block A (such as transcription of the *FOXP3* gene in the nucleus). Figure 5(d) shows that we get fast transient responses for Foxp3 because of the fast transient response of the elements in Block B (i.e. STAT5 and mTORC1), which regulate Foxp3. However, the overall biological behavior of the system is almost the same as in the USB-RandSeq scheme, with faster response since Block B elements are getting updated more often.

### 4.2.5 Ranked-Order Sequential Scheme

In the RankSeq simulation scheme, we can order the rules based on a priori knowledge about the sequence of the biological events. Here, we assign rank 1 for the rules that represent the cell membrane elements (e.g., TCR) and rank 2 for the elements that are regulated by the cell membrane elements, and we continue with the same procedure until we reach the last element in the signaling pathway (e.g., the transcription of a gene has the last rank). Figure 5(e) shows the simulation results for the sub-scenarios of Scenario I using the RankSeq scheme. The results for RankSeq are almost the same as the results shown in Figure 5(b), which were obtained using the RB-RandSeq scheme, suggesting that the RankSeq scheme can be used if the information about the order of the biological events are known. Also, this shows that the RandSeq scheme is able to capture the biological events even if the prior knowledge about the signaling events is not available.

### 4.2.6 Toggling Feature

Here, we run the simulator using USB-RandSeq scheme for the three sub-scenarios II1, II2, and II3, when the TCR signal is turned off at simulation steps 100, 300, and 500, respectively. The simulation results in Figure 5(f) show that the time at which the TCR signal is turned off is critical for the T cell differentiation as confirmed by (Miskov-Zivanov, et al., 2013). It can be seen that turning off the TCR signal at very early simulation step (e.g., Scenario II1) will lead to undifferentiated cells that are characterized by the low expression of both Foxp3 and IL-2. On the other hand, turning off the TCR signal at an intermediate step (e.g., Scenario II2) will cause the naïve T cells to be differentiated into Treg cells that are characterized by high expression of Foxp3. Additionally, turning off the TCR signal at later steps (e.g., Scenario II3) will produce more Th cells which are characterized by low expression of Foxp3. This behavior can be explained by looking at the trajectories of mTORC1 and CD25/STAT5 where the inhibition signal for Foxp3 through mTORC1 lasts longer when we remove the antigen dose at later simulation steps.

## 5 CONCLUSION

System behavior and its response to inputs such as external stimuli or internal perturbations, can be studied using simulations. These studies will help improve understanding of the system, generate hypotheses, or design new experiments. This paper describes the features of our biological system simulator, DiSH. While similar tools have been developed in the past, the contributions of DiSH include simulations of multi-valued model elements, grouping element update functions in several different ways, and the use of delays to simulate biological networks in a more realistic manner.

DiSH is applicable to discrete (including logical) models of complex biological networks. The advantage of discrete models is that they do not necessarily require information about reaction rates and concentrations, which is often not available or impractical to use. Furthermore, the Sim and RandSeq simulation schemes in DiSH enable analysis of both dynamic system behavior and its attractors.

Deterministic discrete model simulations that assume simultaneous element update enable quick attractor analysis. However, when we have prior knowledge about faster and slower events, simultaneous simulations are not a good choice. In the discrete modeling approach, we can incorporate a priori knowledge about observed difference in event rates using delays and probabilistic simulation.

We illustrated in previous sections how different simulation schemes and features of our simulator can emphasize different aspects of the behavior of a biological system. Therefore, our simulator allows scientists to look at biological systems from various perspectives, and learn more about the system by conducting multiple simulations under different conditions.

## 6 ACKNOWLEDGMENTS

This work is supported in part by DARPA award W911NF-14-1-0422.